\documentclass[aps,prd,superscriptaddress,floats,tightenlines,nofootinbib]{revtex4}

\usepackage{bm,graphicx}
\usepackage{dcolumn}
\usepackage{amssymb}

\newcommand{\beq}{\begin{equation}}
\newcommand{\eeq}{\end{equation}}
\newcommand{\p}{\partial}
\newcommand{\conj}{^*}

\begin{document}

\title{Evolution of cosmic superstring networks: a numerical simulation}

\newcommand{\addressSussex}{Department of Physics \&
Astronomy, University of Sussex, Brighton, BN1 9QH, United Kingdom}

\newcommand{\addressTufts}{Institute of Cosmology, Department of
Physics and Astronomy, Tufts University, Medford, MA 02155, USA}

\author{Jon Urrestilla}
\affiliation{\addressSussex}
\affiliation{\addressTufts}

\author{Alexander Vilenkin}
\affiliation{\addressTufts}

\begin{abstract} 

We study the formation and evolution of an interconnected string
network in large-scale field-theory numerical simulations, both in
flat spacetime and in expanding universe.  The network consists of
gauge $U(1)$ strings of two different kinds and their bound states,
arising due to an attractive interaction potential.  We find that the
network shows no tendency to ``freeze'' and appears to approach a
scaling regime, with all characteristic lengths growing linearly with
time.  Bound strings constitute only a small fraction of the total
string length in the network.

\end{abstract} 
 
\maketitle 
 
\section{Introduction} 
 
It has been recently realized \cite{cs} that fundamental strings and 
other string-like objects, such as $D$-strings, can have astronomical 
dimensions and play the role of cosmic 
strings.\footnote{Higher-dimensional $D$-branes with all but one 
dimension compactified will also appear as stringlike objects from a 
macroscopic point of view.}  Observing these objects in the sky would 
provide the most direct test of superstring theory.  Both types of 
string are naturally formed in the course of brane-antibrane 
annihilation at the end of brane inflation 
\cite{MD02,Tye02,Tye03,DV04,CMP04}.  Fundamental ($F$) and $D$-strings 
produced in the aftermath of this annihilation can form $(p,q)$ bound 
states combining $p$ $F$-strings and $q$ $D$-strings.  As a result the 
strings are expected to form an interconnected $FD$-network 
\cite{DV04,CMP04}, with different types of string joined in 3-way 
$Y$-type junctions. 
 
Similar string networks can also be formed in field theory; a simple 
example has been recently given by Saffin \cite{Saffin05}.  His model 
includes two Abelian Higgs models, with an additional coupling between 
the Higgses.  The model has a broken $U(1)_A \times U(1)_B$ symmetry, 
resulting in two types of string, and the coupling is chosen so that 
$A$ and $B$-type strings are attracted to each other and can form 
$(p,q)$ bound states.  An even simpler example is the usual Abelian 
Higgs model.  With a suitable choice of the Higgs and gauge couplings, 
corresponding to the type-I regime, this model allows stable strings 
with arbitrary windings, which can be joined in 3-way junctions 
\cite{DR06}.  Strings with extreme type I properties
can also be formed in models with SUSY flat directions \cite{Morrisey}.
 
String networks with 3-way vertices can also arise in models with
symmetry breaking of the kind $G\to Z_3$ \cite{Aryal86}, as well as in
non-Abelian field theories with several types of strings corresponding
to non-commuting symmetry generators \cite{Kibble76,MG98}. In the
latter case, when two non-commuting strings cross, a third string
starts stretching between them, resulting in two $Y$-junctions.
 
The evolution of cosmic string networks has been a subject of much 
recent discussion and debate.\footnote{One of the key questions is whether or 
not the network gets entangled and ``freezes'', in which case it would 
eventually dominate the energy density of the universe 
\cite{AV84,BS99}.  However, most recent work points to scaling 
evolution.}  Early work on $Z_3$-strings 
\cite{VV87}, using a simple analytic model, suggested 
that at late times the characteristic scale of the network, $\xi = 
(\mu/\rho_s)^{1/2}$, exhibits scaling behaviour, 
\beq 
\xi(t)=\gamma t, 
\label{scaling} 
\eeq 
and 
\beq 
\rho_s/\rho \sim G\mu/\gamma^2. 
\label{rhos} 
\eeq 
Here, $\rho_s$ and $\rho$ are the average energy densities of the 
network and of the universe, respectively, $\mu$  
is the string tension, and $\gamma$ is a constant . A similar model  
was later used for $Z_N$ networks having $N$ strings joined at 
each vertex \cite{CS05}.   
 
The magnitude of $\gamma$ in Eq.~(\ref{scaling}) depends on the rate 
of energy loss by the network.  In $Z_3$ models, the vertices can 
carry an unconfined magnetic charge.  The energy of the network is 
then efficiently dissipated by gauge field radiation from these 
magnetic monopoles.  Another energy loss mechanism is the formation of 
closed loops and of small nets disconnected from the main network.  In 
the absence of magnetic charges, and if loop and net formation turn 
out to be inefficient, the remaining energy loss channel is the 
gravitational radiation.  In this case, the analysis of \cite{VV87} 
gives $\gamma\sim G\mu$, and Eq.~(\ref{rhos}) gives $\rho_s/\rho\sim 
1/G\mu\gg 1$, indicating that the string network becomes so dense that 
it dominates the universe.

More sophisticated analytic models have recently been developed, aimed 
directly to describe cosmic superstrings.  These models allow for 
several types of string with different tensions and use the 
velocity-dependent one-scale model of string evolution 
\cite{Tye05,Shellard07} (see also \cite{Martins04}).  The models make 
somewhat different assumptions about the physics of $F$- and 
$D$-string interaction.  Tye, Wasserman and Wyman \cite{Tye05} assume 
that when $F$ and $D$ strings meet and ``zip'' to form a bound 
$FD$-string, the excess energy is released in the form of high-energy 
particles.  If this picture is correct, it provides an important 
additional mechanism of energy loss by the network.  They also assume 
that the entire network is characterized by a single length scale 
$\xi(t)$.  Avgoustidis and Shellard \cite{Shellard07} assume, on the 
other hand, that the energy released in the zipping process goes to 
increase the kinetic energy of strings, and thus remains in the 
network, and allow different length scales for different string types. 
Assuming that the rate of energy loss to loops from networks is about 
the same as that from ``ordinary'' strings, both models predict 
scaling evolution, with $\gamma\sim 1$, with energy about equally 
distributed between $(1,0)$, $(0,1)$, and $(1,\pm 1)$ 
strings, and with negligible amount of energy is higher-(p,q) strings. 
 
Network evolution has also been studied in field-theory numerical 
simulations.\footnote{Earlier simulations, using a simple model of 
straight strings joined at vertices, were performed in \cite{VV87} for 
a $Z_3$ network and in \cite{MG98} for non-Abelian strings.}  Spergel 
and Pen \cite{SP97} and later Copeland and Saffin \cite{CS05} used a 
non-linear sigma-model to simulate non-commuting string networks. 
Hindmarsh and Saffin \cite{HS06} performed a full field theory 
simulation of global $Z_3$ strings.  In all this work, the network was found 
to scale with $\gamma \sim 0.1 - 1$, indicating efficient damping. 
The models used in \cite{SP97,CS05,HS06} have some important 
differences from superstring networks.  First, all types of string in 
these models have the same tension, while in an $FD$-network the 
tensions of all $(p,q)$ strings are generally different.  Second, the 
global symmetry breaking models used in \cite{SP97,CS05,HS06} allowed 
for an additional energy loss mechanism -- the radiation of massless 
Goldstone bosons -- which is known to be rather efficient.  On the 
other hand, superstring networks are expected to have only 
gravitational-strength couplings to massless (or light)
bosons.\footnote{The possibility of cosmic superstrings having
  stronger than gravitational couplings to massless Ramond-Ramond fields has
  been recently discussed by Firouzjahi \cite{Firouzjahi}.}  Hence, 
there is a danger of string domination, unless the network can 
efficiently lose energy by loop or small net production. 

A field theory simulation of an interconnected string network has
been recently developed by Rajantie, Sakellariadou and Stoica
\cite{RSS}.  They used a model of interacting scalar and gauge fields
similar to the Saffin's model \cite{Saffin05}, which allows two types
of string and a spectrum of bound sates.  The dynamic range of their
simulations was not sufficient to reach any conclusions about the
statistical properties of the network and its scaling behaviour (or
lack thereof).  The main focus of Rajantie {\it et. al.} paper is on
the effect of the long-range interaction induced by the Goldstone
field in models where one of the two $U(1)$ symmetries of the model is
global.  They find that this interaction disrupts the string bound
states in the network.  This result is probably of little relevance
for superstring $FD$-networks, since, as we already noted, superstring
interactions are expected to have gravitational strength and thus have
little effect on network dynamics.

In this paper, we have developed a new network simulation, 
which we believe to be closer to a ``realistic'' superstring network. 
We used a field theory model with two types of gauge strings and 
adopted Saffin's \cite{Saffin05} interaction potential to ensure that 
the strings form bound states.  The details of the model and of the 
simulation are given in the next section.  The results are presented 
in Section 3.  Our conclusions are summarized and discussed in Section 
4.

\section{Simulation details} 
 
\subsection{The model}

Saffin's model of interacting strings\footnote{Similar models
have been studied in relation to composite defects
\cite{BBH03,AU03,B03,AHU05}.} is defined by the Lagrangian
\cite{Saffin05} \beq {\cal L}=|D_\mu\phi|^2 + |{\cal D}_\mu\psi|^2
-{1\over{4}}F_{\mu\nu}^2 -{1\over{4}}{\cal F}_{\mu\nu}^2
-V(|\phi|,|\psi|).
\label{Lagr} 
\eeq 
Here, $\phi$ and $\psi$ are complex scalar fields, charged with
respect to $A_\mu$ and $B_\mu$ gauge fields, respectively,
\beq 
D_\mu =\p_\mu-ieA_\mu, ~~~~ {\cal D}_\mu =\p_\mu-igB_\mu, 
\eeq 
\beq 
F_{\mu\nu}=\p_\mu A_\nu - \p_\nu A_\mu, ~~~~  
{\cal F}_{\mu\nu}=\p_\mu B_\nu - \p_\nu B_\mu, 
\eeq 
\beq 
V(|\phi|,|\psi|)={\lambda_A\over{4}}(|\phi|^2-\eta_A^2)^2 +  
{\lambda_B\over{4}}(|\psi|^2-\eta_B^2)^2  
-\kappa(|\phi|^2-\eta_A^2)(|\psi|^2-\eta_B^2). 
\label{V} 
\eeq 
Without the last term in the potential, the model describes 
independent $A$- and $B$-strings.  Bound states are formed if the 
parameter $\kappa$ is chosen in the range \cite{Saffin05} 
\beq 
0<\kappa<{1\over{2}}\sqrt{\lambda_A \lambda_B}. 
\label{kapparange} 
\eeq 
 
In this paper we shall not attempt to explore the full parameter space 
of the model and consider only the special case where the strings are 
in the Bogomol'nyi limit, 
\beq 
\lambda_A=2e^2, ~~~~ \lambda_B=2g^2. 
\eeq 
We shall also set $e=g$ and $\eta_A=\eta_B$.  With standard 
rescalings, the parameters of the model can then be reduced to 
\beq 
\eta_A=\eta_B=1, 
\eeq 
\beq 
e=g={1\over{2}}\lambda_A={1\over{2}}\lambda_B =1, 
\eeq 
and the range of $\kappa$ in Eq.~(\ref{kapparange}) becomes 
\beq 
0<\kappa<1. 
\label{kapparange2} 
\eeq 
In most of our simulations we used the value
$\kappa=0.9$. Table~\ref{table1} gives the corresponding string
tensions, as well as the binding energies (per unit length of string),
which are relatively large. We have also included in
Table~\ref{table1} the values for a larger $\kappa=0.95$.

\begin{table}[!htb]
\begin{tabular}{|l|cc|cc|}
\hline
      & \multicolumn{2}{c|}{$\kappa=0.90$} & \multicolumn{2}{c|}{$\kappa=0.95$} \\
$(m,n)$ & $\mu$ & $\Delta\mu$ & $\mu$ & $\Delta\mu$ \\\hline
(1,0) & $0.793$ & - ~ & $0.728 $  &  - ~  \\
(1,1) & $1.278$ & $0.308$~  & $1.133$  & $0.323$~  \\
(2,1) & $1.798$ & $0.581$~  & $1.560$  & $0.624$~ \\\hline
\end{tabular}
\caption{String tensions ($\mu$) and bounding energies ($\Delta\mu$) for $(m,n)$ type strings, calculated
for $\kappa=0.90$ \cite{Saffin05} and $\kappa=0.95$. 
}
\label{table1}
\end{table}

\subsection{Numerical setup}
\label{numerics}

Our aim was to perform real-time lattice simulations of model
(\ref{Lagr}), for as long a time as our facilities allowed us. In order
to do this, we obtained the equations of motions from (\ref{Lagr}),
\begin{eqnarray}
 \ddot{\phi} + 2 \frac{\dot{a}}{a}\dot{\phi} - D_{j} D_{j} \phi
 & = &
 -a^{2} \phi \left[\frac{\lambda_A}{2} \left( |\phi|^{2} - \eta_A^{2} \right)
+
\kappa \left(|\psi|^2 - \eta_B^2\right)\right]
\nonumber\\
 \ddot{\psi} + 2 \frac{\dot{a}}{a}\dot{\psi} - {\cal D}_{j} {\cal D}_{j} \psi
 & = &
 -a^{2} \psi \left[\frac{\lambda_B}{2} \left( |\psi|^{2} - \eta_B^{2} \right)
+
\kappa \left(|\phi|^2 - \eta_A^2\right)\right]
\nonumber\\
 \dot{F}_{0j} - \partial_{i}F_{ij}
 & = &
 -2a^{2} e\; \mathcal{I}m \! \left[ \phi\conj D_{j}\phi \right]
\nonumber\\
 \dot{{\cal F}}_{0j} - \partial_{i}{\cal F}_{ij}
 & = &
 -2a^{2} g \; \mathcal{I}m \! \left[ \psi\conj{\cal D}_{j}\psi \right]
 \label{eom}
\end{eqnarray}
where we have made a gauge choice $A_0=B_0=0$ and assumed a flat FRW
spacetime written in conformal time,
\beq
ds^2=a^2(\tau)(d\tau^2-d{\bf x}^2).
\eeq
Overdots in Eqs.~(\ref{eom}) stand for derivatives with respect to
$\tau$.

There is a well-known problem in such simulations: the string core has
a fixed physical width, whereas the distance between lattice-points
grows with the expansion ($a\propto \tau$ in radiation era, and $a\propto
\tau^2$ in matter era). As a result the string width quickly
drops below the resolution threshold of the simulation. Here, we have
adopted the approach used in \cite{Ryden1,Ryden2,moore,BHKU06}, in
which the equations of motion are artificially modified to have the
string width growing with the expansion, so as to be able to simulate them
throughout the evolution.

Following \cite{BHKU06}, the equations of motion (\ref{eom}) can be
written as:
\begin{eqnarray}
 \ddot{\phi} + 2 \frac{\dot{a}}{a}\dot{\phi} - D_{j} D_{j} \phi
 & = &
 -a^{2s} \phi \left[\frac{\lambda_A}{2} \left( |\phi|^{2} - \eta_A^{2} \right)
+
\kappa \left(|\psi|^2 - \eta_B^2\right)\right]
\nonumber\\
\nonumber\\
 \dot{F}_{0j} - 2(1-s)  \frac{\dot{a}}{a} \partial_{i}F_{ij}
 & = &
 -2a^{2s} e\; \mathcal{I}m \! \left[ \phi\conj D_{j}\phi \right]
 \label{eommod}
\end{eqnarray}
and likewise for $\psi$ and $B_\mu$. Here, $s$ is a parameter that
controls the growth of the string width, with $s=1$ being the true
value. As earlier work has shown \cite{BHKU06,moore}, there is little
difference in string dynamics for different values of $s$. For the
remainder of this work, we shall set $s=0$ (that is, the string 
has constant comoving width).

We discretized the modified equations of motion (\ref{eommod}) on a
lattice using the standard lattice link variable approach
\cite{moriarty} and performed the simulations on the UK National
Cosmology Supercomputer \cite{cosmos}. The simulation box consisted of
$512^3$ lattice points, with periodic boundary conditions.  We chose
$\Delta x=1.0$ and $\Delta \tau = 0.2$ trying to maximize the
dynamical range of the simulation.

Scaling evolution regime is expected to be an attractor, and
indeed earlier work has shown \cite{BHKU06,UABL02,PU03,ASU07} that this regime
is approached from a wide range of initial configurations.
Nonetheless, constructing initial conditions for this kind of
simulation is a nontrivial task.  The challenge is to find some
initial configuration that leads to scaling as fast as possible, in
order to maximize the dynamical range.

For the results presented here, we used the following procedure: we
set all gauge fields and gauge field momenta to zero, also set the
scalar field momenta to zero. The magnitudes of the scalar fields is
set to $\eta_A$ and $\eta_B$ respectively (in the present work
$\eta_A=\eta_B=1$), and their phases are chosen randomly.  This
configuration was then smoothed out by averaging over nearest
neighbours, and this procedure repeated 20 times, in order to
get rid of some excess initial gradient energy. Note that this initial
configuration satisfies the (lattice) Gauss law, and due to the
lattice-link variable procedure, the Gauss law is guaranteed to be
satisfied throughout the simulation.

The initial configuration described above corresponds to a highly
excited state, so in order to remove the extra energy from the system,
a fake constant damping was applied, $\gamma= \frac{\dot{a}}{a} =
0.2$, until time $\tau =32$. From then on, we performed simulations
for flat spacetime ($\gamma=0$), radiation ($\gamma=1/\tau$) and
matter ($\gamma=2/\tau$) eras. Only times $\tau>64$ were used for this
analysis.

In order to automatically detect the strings in the
simulation and compute their length, we calculated the net winding of
the phases around plaquettes. One can then trace the string following
the winding and estimate the lengths. The length $L$ of a string that
crosses $n$ plaquettes is estimated as $L=n\cdot\Delta x$.

An $AB$ string can be traced by the sites where both $A$ and $B$
phases wind.  Unfortunately, there are two drawbacks of this
procedure: on the one hand, there are accidental crossings of $A$ and
$B$ strings, i.e., lattice sites where an $A$ string and a $B$ string
simply cross and follow their way, without any attempt to form a bound
state. On the other hand, there are places where inside a clear
segment of $AB$ string, $A$ and B phases do not wind in exactly the
same plaquettes, but there is a slight displacement (we will show an
example in the following Section).  In order to circumvent these
problems, we have excluded accidental crossing by setting a minimum
size of $AB$ string ($l^{AB}_{min}$), and also excluded gaps in
$AB$ segments (due to an occasional ``displacement'') shorter than a
given distance ($d^{AB}_{max}$).  This means that in the process of
detecting $AB$ segments, all segments of size $l<l^{AB}_{min}$ will
not be considered as legitimate segments. Also, if the separation
between two segments is $d<d^{AB}_{max}$, those two segments are
considered as one. The total length of $AB$ string is not considerably
affected by this process; the main difference corresponds to a more
realistic count of the number $N$ of $Y$-junctions formed, that is, it
helps in not overcounting $AB$-segments and $Y$-junctions.  There is
no obvious way of determining the values of $l^{AB}_{min}$ and
$d^{AB}_{max}$, and we have chosen them to be $l^{AB}_{min}=3$ and
$d^{AB}_{max}=5$ by trial and error, and inspection of the results.
With this choice, the corrected value of $N$ is roughly a factor of 4
smaller than one would get from the raw data.

As we already mentioned, our choice of parameters was largely
motivated by the effort to increase the dynamic range of the
simulation. For example, we allocated only a few lattice points per
string thickness.  As a result, our discretized representation of the
field theory string solutions is not particularly accurate.  For a
rapidly moving string, this may result in spurious damping, with the
kinetic energy of the string being dissipated into particles
\cite{Jose}.  Moore {\it et. al.} \cite{moore} performed numerical 
tests to determine the optimal choice of the lattice spacing $\Delta
x$ and concluded that $\Delta x = 0.5$ is close to the maximal value
that still accurately represents the string dynamics.  For larger
values of $\Delta x$, they observed a significant increase in spurious
particle emission from oscillating strings. 

In the present paper, our focus is not so much on the dynamics of
oscillating loops, as it is on the overall characteristics of the large
network. In fact,
in order to observe oscillating loops in field theory simulations
$\Delta x$ should be rather small (even smaller than 0.5), leaving us with a tiny dynamical
range where large network properties would be impossible to study. 
We therefore pushed the parameter values a bit further and
used $\Delta x = 1.0$, with the hope that these properties will not
be strongly affected.  We performed several tests by simulating the
system with different $\Delta x$ ($0.5<\Delta x<1.0$), smoothness of
the initial configuration (smoothing between 0 and 30 times), the
value of the initial fake damping (between 0 and 0.5) and the length
of time in which the fake damping was active ($0<t<50$).  The
qualitative results for all characteristic lengths of the
network were the same in all cases, with the actual values agreeing
within $10-20\%$.  (The only exception is the length of $A$ and $B$
segments, as discussed in Sections~\ref{extradamp} and \ref{true}.)

\begin{figure}[!htb]
\begin{center}
\includegraphics[width=15cm]{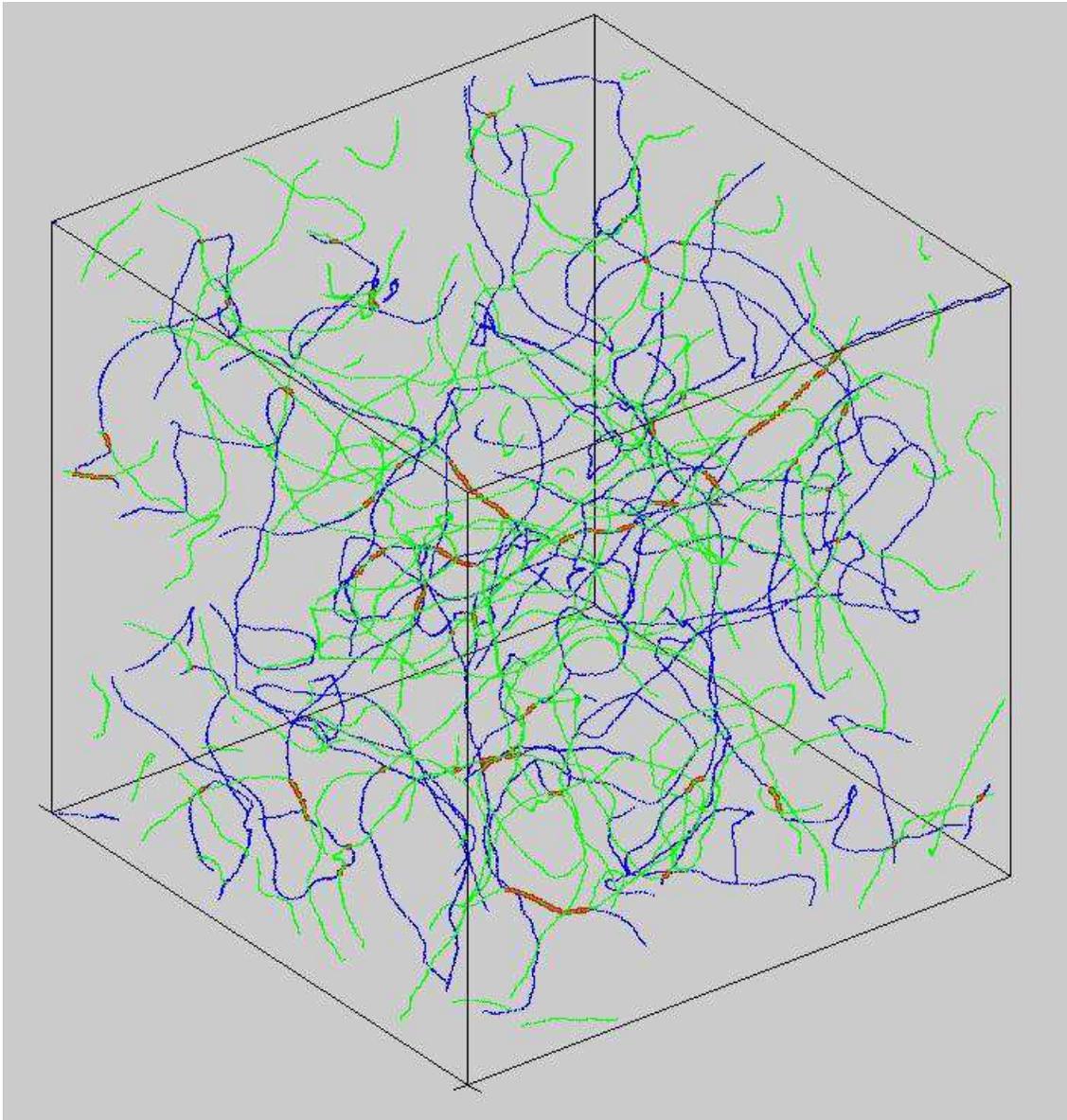}
\caption{\label{both} Picture of a typical simulation of a (p,q) network. 
The green and blue colours correspond to $A$ and $B$ strings
respectively, whereas the red colour shows the $AB$ segments. It can be
clearly seen how Y junctions are formed all over the simulation.}
\end{center}
\end{figure}

\section{Results} 
 
\subsection{The network correlation scale $\xi$}

\begin{figure}[!htb]
\begin{center}
\includegraphics[width=15cm]{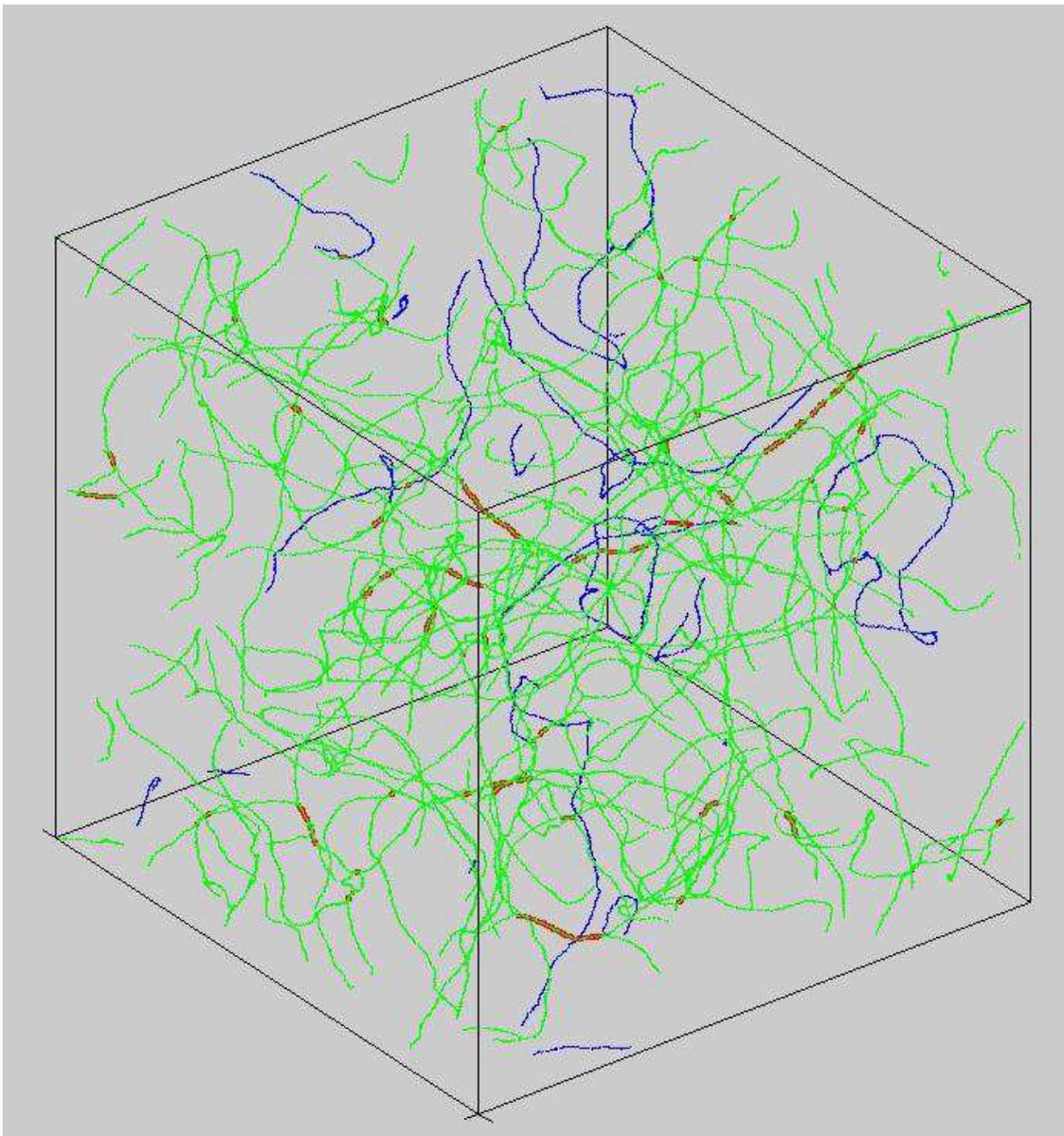}
\caption{\label{ind} Picture of a typical simulation of a (p,q) network. 
This picture represents the same configuration, but we show
how most of the string length (typically more than 90\%) forms an
interconnected network ($A$ and $B$ strings in the main network shown
in green and 
$AB$ segments in red) and there are only a few loops that do not
belong to the main network (blue).  }
\end{center}
\end{figure}

Our simulations were performed in flat spacetime and in the radiation
and matter eras.  In all cases an interconnected network was formed
with $A$-, $B$- and $(1,1)$ $AB$-strings.  No higher-$(p,q)$ strings
were observed.  A representative snapshot of the network is shown in
Fig.~\ref{both}.  Throughout the evolution, the network is
dominated by one large (``infinite'') interconnected net, comprising
 more than 90\% of the total string length, as seen in Fig.~\ref{ind}.

\begin{figure}[!htb]
\begin{center}
\includegraphics[width=8cm]{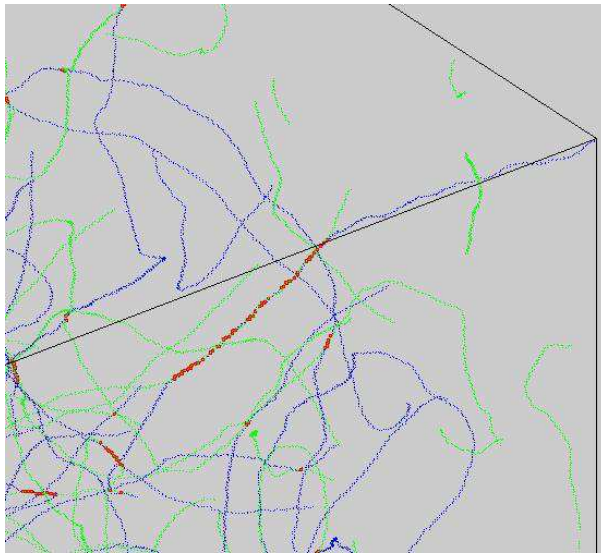}
\caption{\label{displ} Fragment of the simulation box showed in
Fig.~\ref{both}. Green and blue correspond to $A$ and $B$ string,
whereas the red colour corresponds to an $AB$ segment. A long $AB$
segment can be seen in the picture, but at some points the $A$ and $B$
string miss each other by just a lattice point. With the help of the
parameter $d^{AB}_{max}$ those accidental displacements are accounted
for, and long segments such as the one in the figure are counted as
one.}
\end{center}
\end{figure}

As mentioned in the previous section, within a single $AB$ bound
state, $A$ and $B$ strings can be displaced by a single lattice point,
making the code decide that it is in fact two separate
segments. Figure~\ref{displ} shows a fragment of the simulation box,
with a somewhat long $AB$ string depicted. Those accidental
displacements should not be taken into account, and with the help of
the parameter $d^{AB}_{max}$ (introduced earlier), the displacements
are reassessed, and the segment is counted as one.

The overall length scale of the network can be defined as usual,
\beq
\xi = (V/L)^{1/2},
\label{xi}
\eeq
where $V$ is the volume of the simulation box, $L$ is the total
length of string,
\beq
L=L_A+L_B+L_{AB},
\label{L}
\eeq
and $L_A$, $L_B$ and $L_{AB}$ are the lengths in $A$, $B$ and $AB$
strings, respectively.  $\xi$ gives the typical distance between
strings in the network.

Throughout this paper we shall use {\it comoving} length scales.  The
corresponding physical lengths, which will be denoted by superscript
$(ph)$, can be obtained by multiplying with the scale factor
$a(\tau)$, e.g.,
\beq
\xi^{(ph)}(\tau)=a(\tau)\xi(\tau).
\label{xiph}
\eeq
For our models, the scale factor has the form
\beq
a(\tau)=(\tau/\tau_0)^n,
\label{atau}
\eeq
with $\tau_0=const$ and $n=0,1,2$ for flat spacetime, radiation and matter
eras, respectively.  The physical time $t$ and the horizon distance
$\ell_H$ are given by
\beq
t=\int_0^\tau a(\tau')d\tau' = {\tau^{n+1}\over{(n+1)\tau_0^n}},
\label{t}
\eeq
\beq
\ell_H=a(\tau)\tau =(n+1)t.
\label{horizon}
\eeq

\begin{figure}[!htb]
\begin{center}
\includegraphics[width=8cm]{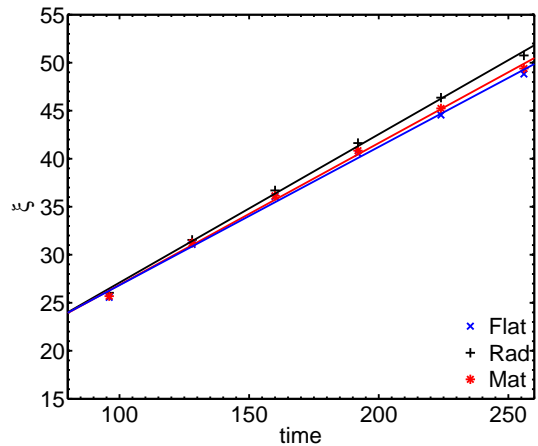}
\caption{\label{xiplot} The correlation length $\xi$, as defined
in Eqn.~\ref{xi}, averaged over 10 simulations for flat (dashed
black), radiation (continuous red) and matter (dotted blue)
regimes. Note that all three cases exhibit an approximately
linear behaviour, nearly independent of the regime.  }
\end{center}
\end{figure}

The simulation results for $\xi(\tau)$ are shown in Fig.~\ref{xiplot},
where each graph represents an average over 10 simulations.
Remarkably, the graphs for the flat, radiation and matter regimes are
almost identical. They show a nearly linear dependence,
\beq
\xi(\tau)=\alpha\tau+\xi^{(0)},
\label{linear}
\eeq
with $\alpha\approx 0.15$ (see Table~\ref{table2}).  Toward the end of
the simulation, the two terms in Eq.~(\ref{linear}) are comparable to
one another.  If the linear dependence extends to much larger values
of $\tau$, the constant term eventually becomes negligible,
\beq
\xi\approx\alpha\tau,
\label{xitau}
\eeq
and the corresponding physical length grows proportionally to the
horizon,
\beq
\xi^{(ph)}(t)\approx (n+1)\alpha t =\alpha\ell_H(t).
\label{xiphH}
\eeq

\begin{table}[!htb]
\begin{tabular}{|l|c|c|c|}
\hline
      & Flat & Radiation & Matter \\ \hline
$\alpha$ ($\xi$) & 0.14 & 0.15 & 0.15 \\
$\alpha_A$ ($\xi_A$) & 0.21 & 0.22 & 0.21 \\
$\alpha_{AB}$ ($l_{AB}$) & 0.03 & 0.07 & 0.08 \\
$\alpha_N$ ($\xi_N$)  & 0.21 & 0.28 & 0.30\\\hline
\end{tabular}
\caption{The values of the linear growth coefficients of different 
lengths, as defined in the body of the paper, obtained by fitting the
simulation data. Note that with our parameter values $\alpha_A=\alpha_B$.}
\label{table2}
\end{table}

\subsection{Bound strings}
\label{bound}

An important characteristic of the network is the fraction of
total string length in bound ($AB$) strings, 
\beq
f_{AB}={L_{AB}\over{L_A+L_B+L_{AB}}}.
\label{fAB}
\eeq 

\begin{figure}[!htb]
\begin{center}
\includegraphics[width=8cm]{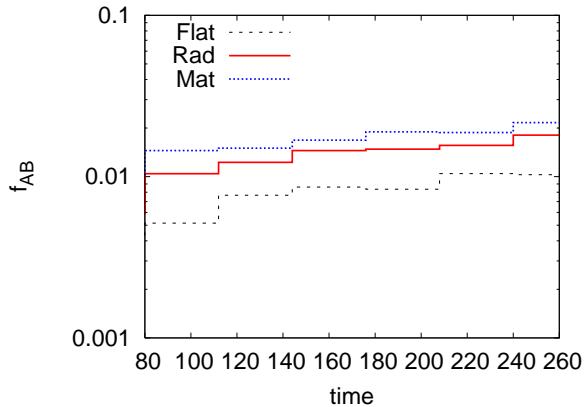}
\caption{\label{boundf} Fraction of total string length in bound strings, 
for flat, radiation and matter regimes. The percentage is fairly low;
between 0.5-1$\%$ for flat, and 1-2$\%$ for radiation and matter
cases.  }
\end{center}
\end{figure}

The simulation results for $f_{AB}$ are shown in Fig.~\ref{boundf}. We see that
$f_{AB}$ remains nearly constant, at the value $0.01\leq f_{AB}\leq0.02$ in
radiation and matter eras, and somewhat lower for flat spacetime.
Hence, bound strings constitute less than 2\% of the network.  This is
in conflict with analytic models \cite{Tye05,Shellard07} predicting
that the energy of the network should be more or less equally divided
between $A$, $B$ and $AB$-strings.  The main source of the discrepancy
is the assumption made in \cite{Tye05,Shellard07} that crossings of
$A$ and $B$ strings typically lead to the formation of relatively
stable bound $AB$ segments of length $\sim\xi(t)$.  Visual inspection
of the simulation movies suggests, on the contrary, that formation of
bound segments by intersecting $A$ and $B$ strings occurs rather
infrequently, probably when the relative velocity of the colliding
strings is sufficiently small \cite{CKS06,CKS07}.  Even when they are formed, the
$AB$-segments easily ``unzip'' as the free $A$ and $B$ ends pull in
different directions and do not usually last for more than a Hubble
time.

\begin{figure}[!htb]
\begin{center}
\includegraphics[width=8cm]{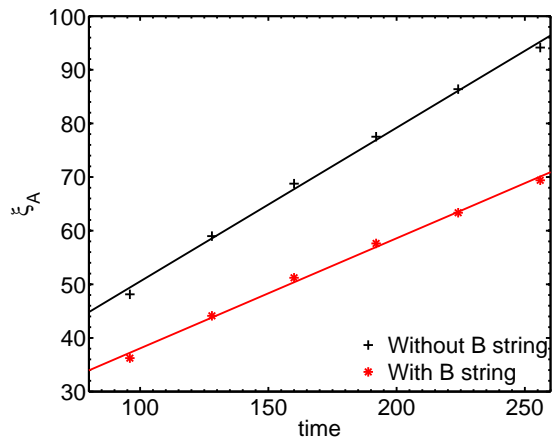}
\caption{\label{u1} Comparison of $\xi_A$ (only in radiation era, for clarity)
between simulations points and without $B$-strings.  }
\end{center}
\end{figure}

Even though the fraction of bound string is small, the interaction of
$A$ and $B$-strings has a significant effect on the network evolution.
To quantify this effect, we ran some simulations with the same initial
conditions for $A$-string fields as before, but with $B$-string fields
starting in their ground state, so that no $B$ strings are formed.
$A$-strings then evolve as ordinary $U(1)$ strings, and their
characteristic length 
\beq
\xi_A=(V/L_A)^{1/2}
\label{xiA}
\eeq
exhibits a linear dependence on $\tau$ with $\alpha_A\approx 0.29$, as
seen in Fig.~\ref{u1}.  (This value is in agreement with earlier
$U(1)$ simulations by Vincent et. al. \cite{Vincent} and Moore
et. al. \cite{moore}.)  The same quantity calculated with $B$-strings
present gives $\alpha_A\approx 0.22$, so the growth of $\xi_A$ is
significantly slower than it would be if the two kinds of string
evolved independently.

\begin{figure}[!htb]
\begin{center}
\includegraphics[width=8cm]{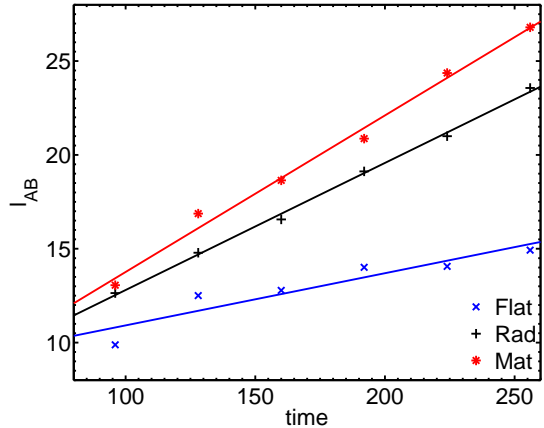}
\caption{\label{lab} The average length of bound segments $l_{AB}$, 
as defined in Eq.~\ref{lAB}, for flat, radiation and matter regimes.
The evolution is approximately linear, with slope given in
Table~\ref{table2}.  }
\end{center}
\end{figure}

The average (comoving) length of $AB$-segments is
\beq
l_{AB} = L_{AB}/N,
\label{lAB}
\eeq
where $N$ is the number of $AB$-segments ($2N$ is the number of
$Y$-junctions where the three types of string meet).  Fig.~\ref{lab}
shows that the evolution of $l_{AB}$ is approximately linear,
\beq
l_{AB}\approx\alpha_{AB} \tau + l_{AB}^{(0)}.  
\label{lABtau}
\eeq

The coefficient $\alpha_{AB}$ (see Table~\ref{table2}) is nearly the
same in radiation and matter eras (within 10\%),
$\alpha_{AB}^{(rad,mat)}\approx 0.07$, and is significantly smaller in
flat spacetime, $\alpha_{AB}^{(flat)}
\approx 0.025$.  The shorter bound segments in flat spacetime are probably 
due to larger string velocities.

\begin{figure}[!htb]
\begin{center}
\includegraphics[width=8cm]{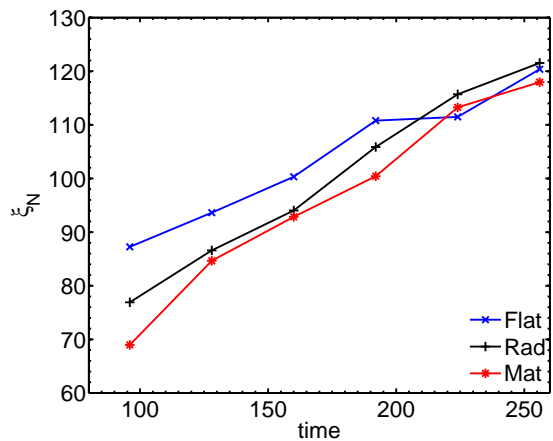}
\caption{\label{xin} The average distance between $Y$-junctions,
$\xi_N$ as defined in Eq.~\ref{xiN}, for flat, radiation and matter
regimes.  The evolution is approximately linear, with slope given in
Table~\ref{table2}.  }
\end{center}
\end{figure}

The typical distance between $Y$-junctions is given by
\beq
\xi_N=(V/2N)^{1/3}.
\label{xiN}
\eeq
Once again, we find approximately linear evolution
(see Fig.~\ref{xin}),
\beq
\xi_N\approx \alpha_N \tau +\xi_N^{(0)},
\eeq
with $\alpha_N\approx 0.3$.

\subsection{Non-scaling of $A$ and $B$ segments}
\label{extradamp}

\begin{figure}[!htb]
\begin{center}
\includegraphics[width=8cm]{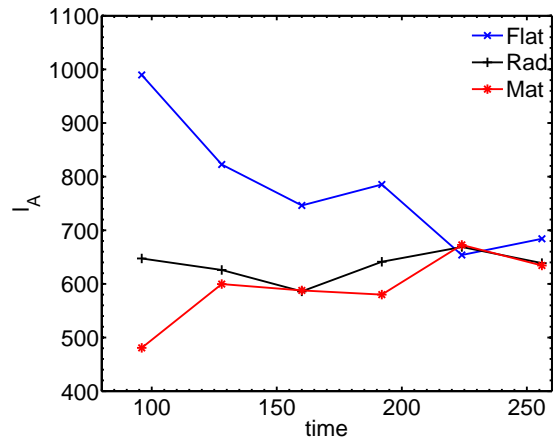}
\caption{\label{la} The average length of $A$-segments $l_A$ is
shown for flat, radiation and matter regimes.  In all three cases,
$l_A$ is nearly constant at late times. The behaviour of $l_B$ is
essentially the same.}
\end{center}
\end{figure}

All results presented so far are consistent with scaling evolution,
with all characteristic length scales of the network growing
proportionally to the horizon.  However, the average comoving lengths
of $A$ and $B$ segments,
\beq
l_A = L_A/N, ~~~~~ l_B = L_B/N,
\label{xiAB}
\eeq
do not exhibit scaling behaviour.  In fact, Fig.~\ref{la} shows that, rather
surprisingly, these lengths approach nearly constant values, 
\beq
l_A\approx l_B \approx {\rm const}.
\label{C}
\eeq
The corresponding physical lengths grow proportionally to the scale
factor.  

Throughout the simulation, the segment length $l_A$ remains much
greater than the correlation length $\xi_A$.  The correlation length
grows with time, but is still about an order of magnitude smaller than
$l_A$ at the end of the simulation (and similarly for $l_B$ and
$\xi_B$).  This suggests that $A$ (and $B$) segments have the shape of
random walks of step $\sim\xi_A$ and end-to-end distance $\sim \xi_N$.
The length of the segments is then 
\beq
l_A\sim \xi_N^2/\xi_A. 
\label{laxin}
\eeq
We have verified that the ratio $\xi_N^2/\xi_A l_A =\xi_A/2\xi_N$
is indeed approximately a constant of order 0.3, within $10-20\%$.

If the length of the segments $l_A$ continues to grow slower than
their end-to-end distance $\xi_N$, the two lengths will eventually
become comparable to one another, with segments becoming more or less
straight.  This could mark the beginning of the true scaling regime,
where all the characteristic lengths of the network have the same
order of magnitude and grow proportionally to $\tau$.  In any case,
the evolution laws (\ref{linear}),(\ref{lABtau}),(\ref{C}) cannot
continue indefinitely and must stop at some $\tau=\tau_*$.  
The situation here may be somewhat similar to that with ``ordinary'' (not
interconnected) strings, where scaling of the characteristic length 
$\xi(t)$ is quickly established, but scaling of the small wiggles on long 
strings and of closed loops is reached only after a long transient period
\cite{OVV06,Shellard06,Sakellariadou07,OV06}.  We shall refer to the 
evolution at $\tau < \tau_*$ as the transient scaling regime.

\subsection{Towards a true scaling regime}
\label{true}

We attempted to shorten this transient regime, or avoid it altogether,
by increasing the duration of the initial damped period.  This has the
effect of increasing $N$ and decreasing the ratio $l_A/\xi_N \sim
(\xi_N/\xi_A)^2$ in the initial state (right after damping is turned
off). Getting to the end of the transient regime by the end of the
simulation proved to be a difficult task, for which we had to
push the limits of the stability of the code by using a rather large
time step, a rather large damping coefficient, and evolving the system
beyond the half-light-crossing time of the simulation box.

We performed a set of 20 simulations with the values of time step
$\Delta t$ and the damping parameter $\gamma$ shown in
Table~\ref{extra}.  The difference from the rest of our simulations is
in the second period of damping imposed after the first period (which
is common to all simulations and is used to relax the system from the
highly excited initial state and allow string formation).  The extra
damping period has a time-discretization of $\Delta t=0.5$ (keeping
$\Delta x=1$), and the numerical stability is achieved because the
damping is also rather high ($\gamma=2.0$).

\begin{table}[t]
\begin{tabular}{|c|c|c|}
\hline
Physical time & Damping ($\gamma$) & $\Delta t$\\
\hline
$0<t<32$ &  0.5 & 0.2 \\
$32<t<272$ & 2.0 & 0.5 \\
$t>272$ & $2.0/t$ (rad.) & 0.2 \\\hline
\end{tabular}
\caption{Values of the parameters in simulation with a second
damping period, introduced to achieve scaling of the segment lengths
$l_A$ and $l_B$.  }
\label{extra}
\end{table}

\begin{figure}[!htb]
\begin{center}
\includegraphics[width=8cm]{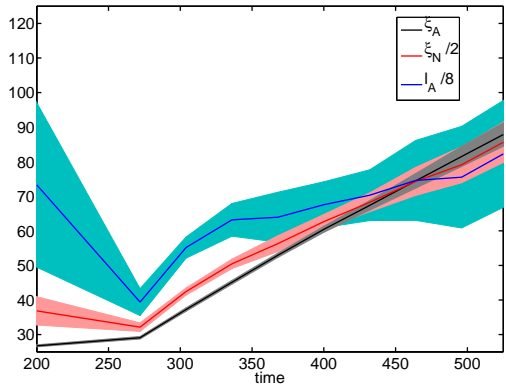}
\caption{\label{extrad} The length measures $\xi_A$, $\xi_N$, 
$l_A$ averaged over 20 simulations performed with an extended damping
period, as explained in Section~\ref{extradamp}.  The thick lines
corresponds to the average, whereas the shaded regions correspond to
1-$\sigma$ statistical errors over the simulations.  }
\end{center}
\end{figure}

Figure~\ref{extrad} shows the length measures $\xi_A$, $\xi_N$, $l_A$
averaged over the 20 simulations.
All three lines seem to approach
a linear behaviour.  The slopes of $\xi_A$ and $\xi_N$ agree with the
ones obtained earlier, without an extended damping period. Other
quantities characterizing the system, such as $\alpha_{AB}$, $f_{AB}$
and the percentage of string length in loops, also agreed with the ones
for the simulations without extra damping. The fairly linear
behaviour of $l_A$ is the main change. A linear fit to the evolution
of $l_A$ gives
\beq
l_A\approx \alpha_{LA}\tau + l_A^{(0)},
\eeq
with $\alpha_{LA} \approx 0.8 $.  Extrapolating this behaviour to
large $\tau$, we expect a scaling regime in which $l_A$ is a few
times larger than $\xi$ and $\xi_N$.  However, because of our extreme
choice of parameters and large error bars in $l_A$, and because
the change in $l_A$ during the linear regime is relatively small,
this conclusion should be regarded as tentative.

\subsection{Loops and small nets}

\begin{figure}[!htb]
\begin{center}
\includegraphics[width=8cm]{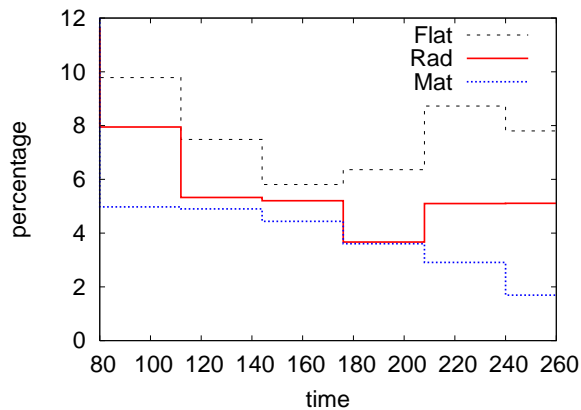}
\caption{\label{loops} Percentage of string length for both $A$ and $B$
strings in loops that do not belong to the main network. 
}\end{center}
\end{figure}

We observed the formation of small independent nets a few times in the
course of the simulation, but these occasions were rather rare, so
$AB$-strings belonged predominantly to the infinite network, with at
most one small net in addition.  The fraction of the total string
length in disconnected $A$- or $B$-loops is shown in Fig.~\ref{loops}.
We see that this fraction remains nearly constant, at a value $f_L\sim
0.03 - 0.05$ in radiation and matter eras and $f_L\sim 0.08$ in flat
spacetime.

\begin{figure}[!htb]
\begin{center}
\includegraphics[width=8cm]{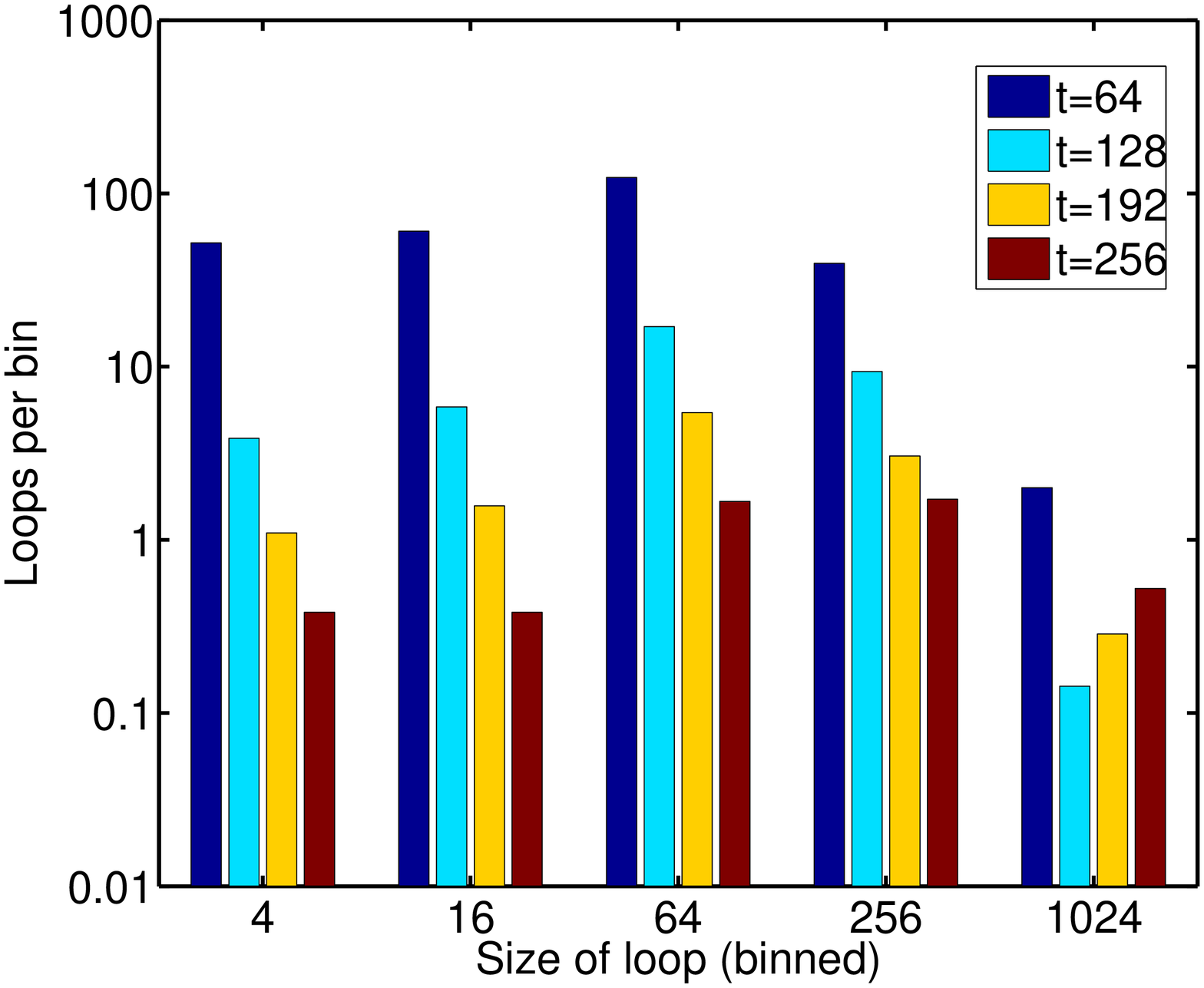}\includegraphics[width=8cm]{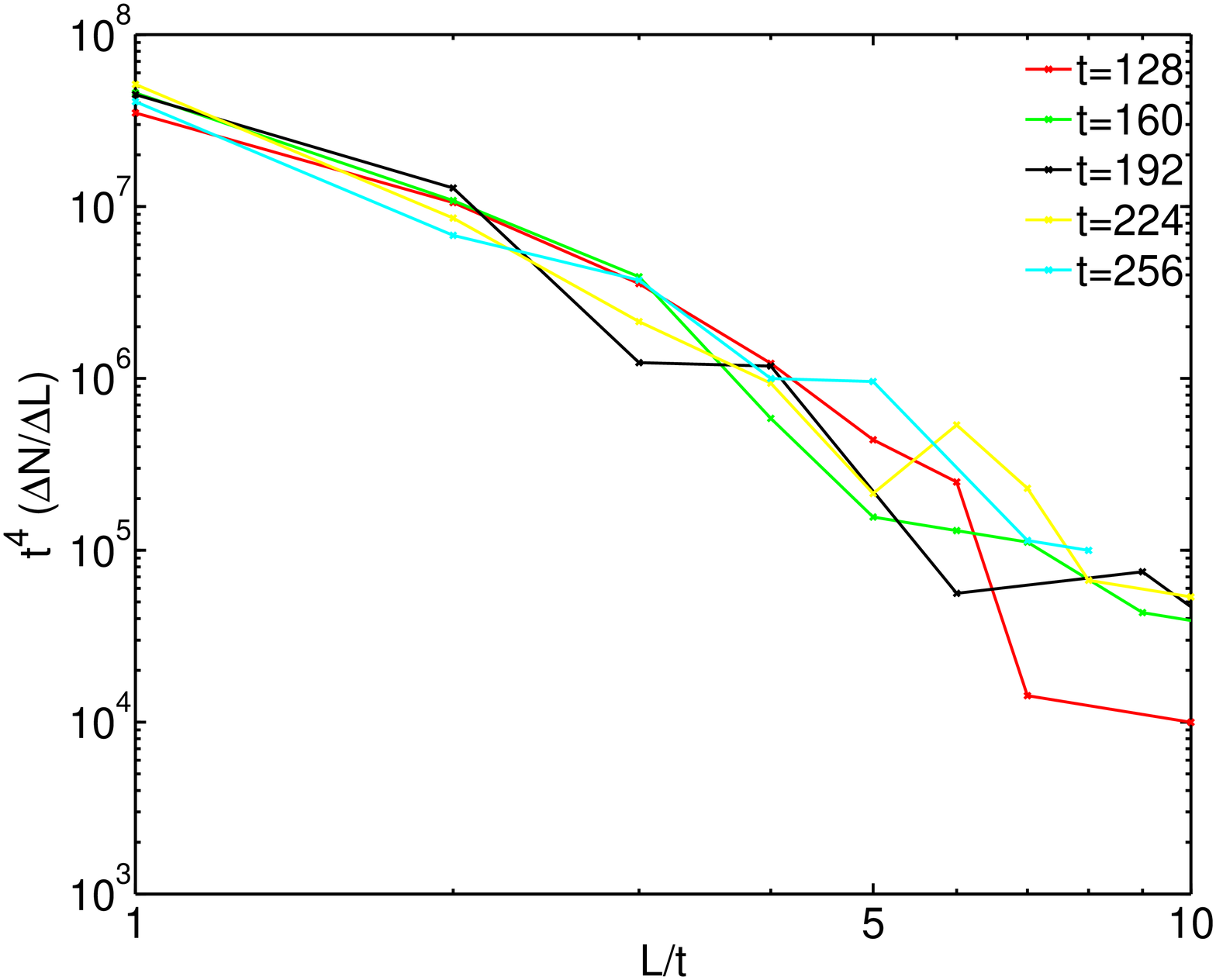}
\caption{\label{loopsb} (Left) Distribution of independent loops for the 
radiation era, averaged over 20 simulations, for four different time
steps. The figure shows that there are rather large independent loops
formed during the simulation, and that the peak of the distribution
tends to move towards larger sizes with time. (Right)
Loop distribution averaged over 20 simulations
in radiation era. The distribution appears to
follow Eqn.~\ref{ntl} rather well, suggesting that it is
scaling.}
\end{center}
\end{figure}

The length distribution of independent loops is plotted in
Fig.~\ref{loopsb}, averaged over 20 simulations in radiation era. The
figure uses logarithmic binning and shows the distribution of loops at
four different times.  We see that some of the loops are fairly large,
with length much greater than the inter-string separation $\xi$.  Such
loops should arguably be regarded as part of the infinite network.
Indeed, if the diameter of a loop is larger than a few times $\xi_A$
(or $\xi_B$) for a given time, the loop is very likely to reconnect to
the network. If we exclude loops longer than $6\xi_A$ ($3\xi_A$), the
percentage of string length in loops drops to $1-2\%$ ($0.5\%$). These
values are in agreement with previous field theoretical simulations
where a loops were found to account for a few percent of the total
string length \cite{Vincent}.

Simulation movies show that loops that decouple from the network do
not oscillate as they would in Nambu-Goto simulations, but rather
shrink and disintegrate.  This could be expected, since it is well
known that in order to observe oscillating loops in field theory
simulations one would need very large loops and very small values of
the lattice spacing.  Additional damping, through particle emission
from loops, may be due to the presence of short-wavelength string
excitations in the initial conditions, as indicated by the numerical
results of Refs.~\cite{Moore98,Olum00}.  The short lifetimes of the
loops in our simulations explain, at least in part, the relatively
small amount of string in loops, as compared to the Nambu-Goto
simulations.

In any case, we checked the loop distribution in our simulations
for scaling behavior.  On dimensional grounds, the number density of
loops per unit length interval in a scaling network should have the form 
\beq
{dn\over{dl}}(t)=t^{-4}f(l/t).
\label{ntl}
\eeq
Defining $\Delta N_i$ as the number of loops between sizes $l_i$ and
$l_{i+1}$ and $\Delta l_i=l_{i+1}-l_i$, we plot in Fig.~\ref{loopsb}
the (binned) quantity $\frac{\Delta N_i}{\Delta l_i} t^4$, obtained
for 5 different times and averaged over 20 simulations. In a situation
where the loop distribution scales, all the lines should line on top
of one another.  We see that the graph in Fig.~\ref{loopsb} does
indeed exhibit scaling.

\subsection{Effect of higher binding energy} 

\begin{figure}[!htb]
\begin{center}
\includegraphics[width=8cm]{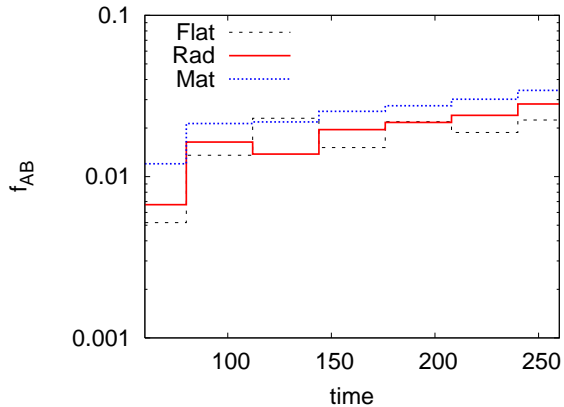}
\caption{\label{bound95} Fraction of total string length in bound strings, 
for flat, radiation and matter regimes, for simulations with
$\kappa=0.95$. The percentage is still fairly low, though somewhat
higher than for $\kappa=0.90$ (Fig.~\ref{boundf})}
\end{center}
\end{figure}

We ran some simulations with a greater value of the string binding
parameter, $\kappa=0.95$, in order to try to understand how sensitive
our results were to the precise value of $\kappa$.  The main
difference encountered in these simulations was the fraction of bound
string $f_{AB}$ increased by a factor $\sim 1.5$ (see
Fig.~\ref{bound95}).  Our analysis showed that the effect on the
correlation lengths $\xi$ and on the loop distribution was very
small. Actually, the values of the coefficient of the different length
measures where the same as in the $\kappa=0.9$ case, the only
difference being a smaller $\alpha_N$ for this case ($0.16,0.22,0.25$
for flat, radiation and matter eras, respectively).

\section{Conclusions and discussion} 
 
We performed numerical simulations of the formation and evolution of 
string networks using the Saffin's model for interacting strings.  The 
model has two types of Abelian gauge strings, $A$ and $B$, with an 
attractive interaction which gives rise to bound $AB$-strings. 
Starting with a randomized high-energy field distribution, we found 
that an interconnected string network is indeed formed, consisting of 
$A$ and $B$ strings, as well as their $(1,1)$ bound states, joined 
together at $Y$-type junctions.  No higher $(p,q)$-strings were 
observed in the simulations.  
 
Throughout each simulation, the network is dominated by a single
(``infinite'') net, with occasional small nets and a fair number
of disconnected closed loops being formed in the course of the
evolution.  The characteristic length scale of the network approaches
the scaling regime where it grows proportionally to time,
$\xi(\tau)=\alpha \tau$ with $\alpha\sim 0.15$, in both radiation and
matter eras, as well as in flat spacetime. Other characteristic length
scales, such as the length of $AB$ segments, the typical correlation
length of $A$ and $B$ strings or the typical distance between $Y$
junctions also scale.

A surprising feature of our simulation is that bound
$AB$-strings constitute only a small fraction of the total string
length ($\sim 2\%$).  Also, the average length of $AB$-segments is
much shorter than the length of $A$- or $B$-string segments.  This is
in contrast with analytic models \cite{Tye05,Shellard07} predicting
all lengths to be fairly equal.  From movies of the simulations one
can see that $AB$-segments do not always form when an $A$-string meets
a $B$-string (even for relatively high bounding energies); on the
contrary, the formation of bound segments is rather infrequent. Even
though new $AB$-segments are constantly formed, their lifetime is
relatively short, usually less than the Hubble time, as the segments
are ``unzipped'' by the free $A$ and $B$ ends pull them in different
directions.

Even though, with our general initial condition configuration, most of
the typical distances in the network showed a scaling behaviour, the
average comoving lengths of $A$ and $B$ segments did not. These
lengths remained much larger than the typical correlation length
throughout the simulation. This evolution regime cannot continue
indefinitely, since the correlation length $\xi$ would eventually
catch up with the segment lengths $l_A$ and $l_B$.  In an attempt to go
beyond this transient regime and reach true scaling, we introduced an
extra period of strong damping at early times (which has the effect of
decreasing $l_A/\xi$) and pushed the parameters to extend the dynamic
range of the simulation.  We found that the system does seem to
approach a regime where all characteristic lengths scale linearly with
time.  If this were the true scaling regime, it would provide us,
among other things, with the means to calculate CMB power spectra
predictions from the field theoretical model as in
\cite{BHKU06,Bevis:2007gh,Bevis:2007qz,Urrestilla:2007sf}. 
But due to a relatively short duration of the linear evolution and to
likely presence of spurious damping, our simulations cannot be relied
upon for a quantitatively accurate description of scaling.

Disconnected closed loops constitute about $10\%$ of the total string
length.  Some of these loops are very large and will very likely
reconnect into the main network again.  If one factors out
disconnected loops of length larger than a few times the
correlation length, the total string length in loops drops to below
$2\%$,  in agreement with \cite{Vincent}.  We examined the length distribution of loops in the network
and found that, even though out parameter choice is not expected to resolve
accurately string dynamics, these distributions seem to scale. 
 
The network properties in our simulations are closer to those of
superstring networks than they were in earlier simulations that used
$Z_3$-strings or non-linear sigma-models.  However, there is still an
important differences.  On the one hand, our $A$ and $B$ strings have
the same tension, as opposed to $F-$ and $D-$ strings. On the other,
collision of same-type strings in our model always result in
reconnection (unless the string segments are moving extremely
 fast \cite{AP06}), while in the case of superstrings the reconnection
probability is $p<1$ and can even be small
\cite{Tye03,DV04,JJP05}. This feature can be accounted for in
Nambu-Goto and analytic models.

The efficiency of various energy loss mechanisms by the string
network remains a topic for future research.  Energy loss to loop
production appears to be substantial, considering that the length in
loops at any time is a few percent of the total and that the loops do
not stay around for long and rapidly decay.  Another important energy
loss mechanism in field theory simulations is direct particle emission
from strings \cite{Vincent,BH07}.  In fact, the analysis in
\cite{Shellard07} shows that emission of particles and of tiny loops
which immediately decay into particles is the dominant energy loss
mechanism for a single $U(1)$ string network, so it probably dominates
in our simulations as well.  It has been argued in
\cite{Moore98,Olum00,Shellard07} that this effect is spurious and is 
due partly to insufficient resolution of the simulations and partly to
excessive amount of noise in the initial conditions.  This issue is
not completely settled, since some of the recent Nambu-Goto
simulations \cite{Sakellariadou07} and analytic treatments
\cite{Polchinski07} indicate continuous production of microscopic
loops throughout the network evolution.

In summary, what have we learnt from our simulations?  We have
demonstrated that an interconnected network of strings can indeed form
at a symmetry breaking phase transition.  This network shows no
tendency to freeze to a static configuration.  On the contrary, it
appears to approach scaling, with all characteristic lengths growing
linearly with time.  Qualitatively, our results indicate that bound
strings constitute only a small part of the total string length and
that the $A$ and $B$ string segments are rather wiggly, having lengths
significantly greater than the correlation length $\xi$.  The latter
property leads to relatively frequent self-intersections and allows
the network to lose a substantial fraction of its energy in the form
of closed loops.  Our simulations also indicate that the true scaling
evolution may be preceded by a transient regime in which the comoving
lengths of $A$ and $B$-segments remain nearly constant in time.

Some of the shortcomings of our approach can be overcome
in Nambu-Goto-type simulations (e.g., of the kind developed 
in \cite{Siemens01}) or in analytic models (along the lines of 
\cite{Tye05,Shellard07}).  Either of these approaches, however, 
requires some microphysical input.  For example, one needs to know 
under what conditions a bound string is formed in a string collision, 
what fraction of the binding energy of the newly formed string is 
radiated away, etc.  The advantage of a direct field theory simulation 
is that it accurately represents the microphysics.  A combination of 
all three approaches will probably be needed to reach a full
understanding of network evolution.

\section{Acknowledgements} 
 
We would like to thank Jos\'e Juan Blanco-Pillado, Mark Hindmarsh, 
Andrew R. Liddle, Ken
Olum and Vitaly Vanchurin for useful discussions. This work was
supported by the US National Science Foundation (JU and AV), by Marie
Curie Intra-European Fellowship MFIT-CT-2005-009628 and FPA2005-04823
(JU). This work has also been supported by
the Spanish Consolider-Ingenio 2010 Programme CPAN (CSD2007-00042).The simulations were performed on cosmos, the UK National Cosmology Supercomputer, 
supported by SGI, Intel, HEFCE and PPARC.

\end{document}